\documentclass[12pt]{article}
\usepackage[dvips]{graphicx}
\usepackage{amssymb}
\usepackage{amsmath,amsthm}

\usepackage{bm}
\bmdefine{\Bx}{x}
\bmdefine{\By}{y}
\bmdefine{\Bone}{1}
\bmdefine{\Bbeta}{\beta}
\bmdefine{\Bvarepsilon}{\varepsilon}
\bmdefine{\Bzero}{0}
\newtheorem{definition}{Definition}[section]
\newtheorem{theorem}{Theorem}[section]
\newtheorem{proposition}{Proposition}[section]

\title{Some optimal criteria of model-robustness for two-level non-regular
fractional factorial designs
}
\author{Satoshi Aoki\footnote{Graduate School of Science and
Engineering, Kagoshima University}\footnote{JST, CREST}
}

\date{November, 2009}

\begin{document}

\maketitle

\begin{abstract}
We present some optimal criteria to evaluate model-robustness of
 non-regular two-level fractional factorial designs. Our method is based
 on minimizing the sum
 of squares of all the off-diagonal elements in the information
 matrix, and considering expectation under appropriate distribution functions
 for unknown contamination of the interaction effects. 
By considering uniform distributions on the symmetric support, our criteria
 can be expressed as linear combinations of $B_s(d)$ characteristic,
 which is used to characterize the generalized minimum aberration. We
 give some empirical studies for $12$-run non-regular designs to evaluate
 our method.\\
Keywords{non-regular designs \and fractional factorial designs \and
 robustness \and affinely full-dimensional factorial designs \and
 $D$-optimality}
\end{abstract}

\section{Introduction}
The most commonly used designs for two-level factorial experiments are
regular fractional factorial designs. This is because properly chosen
regular fractional factorial designs have many desirable
properties such as being orthogonal and balanced. In addition, 
we can easily consider important concepts such as resolution and
aberration for the regular fractional factorial designs in applications. 
For example, under the hierarchical assumption, i. e, lower-order effects
are more important than higher-order effects and effects of the
same order are equally important, a minimum aberration
criterion by Fries and Hunter (1980) seems natural and widely used.
As another reason for using regular designs, an elegant theory
based on the linear algebra over $\mathbb{F}_2$ is 
well established for regular two-level fractional factorial designs. See
Mukerjee and Wu (2006) for example. The only drawback of using regular
fractional factorial designs is that its run size must be a power of
$2$. Therefore if the run size of the design is restricted not to be a
power of $2$ by some cost or manufacturing limitations, we must consider
non-regular fractional factorial designs. See Xu, Phoa and Wong (2009) 
for recent developments in non-regular fractional factorial designs.

One approach of optimal selection for non-regular designs is various
extension of the minimum aberration criterion to non-regular designs.
For example, Deng and Tang (1999) proposed a generalized minimum
aberration criterion, which is a natural extension of the minimum
aberration criterion from regular to non-regular designs. Tang and Deng
(1999) also proposed a minimum $G_2$ aberration criterion, which is
a simpler version of the generalized minimum aberration criterion. 
To justify these criteria, one approach is to evaluate these
criteria from the viewpoint of model-robustness. For example, 
Cheng, Steinberg and Sun (1999) shows that the designs with the minimum
aberration have a good property of model-robustness. Similarly, 
Cheng, Deng and Tang (2002) also investigate the generalized minimum
aberration criterion from the viewpoint of model-robustness.
In this paper, we follow these works and give a new criterion for
model-robustness. Our new criterion is obtained as an extension of the
approach by Cheng, Deng and Tang (2002). We consider contamination of
two- and three-factor
interaction effects for estimating the main effects, whereas Cheng,
Deng and Tang (2002) only considers contamination of the two-factor
interaction effects.

Another approach of choosing non-regular fractional factorial designs is
proposed recently by Aoki and Takemura (2009). Aoki and Takemura (2009)
defines a new class of two-level non-regular fractional factorial
designs, called an affinely full-dimensional factorial design. 
The design points in the design of this class are not contained in any
affine hyperplane in the vector space over $\mathbb{F}_2$. Aoki and
Takemura (2009) also investigates the property of this class from the
viewpoint of $D$-optimality. However, the arguments of Aoki and
Takemura (2009) is restricted to the models of the main effects, and the
property of this class in the case of the presence of the interaction
effects is not yet obtained. In this paper, we also investigate the
relation between our new criteria and the affinely full-dimensional
factorial designs.

The construction of this paper is as follows. In Section
\ref{sec:preliminaries}, we give necessary definitions and notations for
our criteria. We also review the generalized minimum aberration
criterion and the affinely full-dimensional factorial designs briefly.
In Section \ref{sec:optimal-criteria}, we give definitions of our
optimal criteria. One of the important contributions of this paper is to
show the relation between our criteria and the generalized minimal
aberration criterion. For this point, we give a general method to handle
this problem and evaluate values for some cases. We also give 
empirical studies for $12$-run non-regular designs.

\section{Preliminaries}
\label{sec:preliminaries}
First we give necessary definitions and notations for our criteria.
We use some of notations by Cheng, Deng
and Tang (2002). We also review the 
generalized minimum aberration criterion and affinely full-dimensional
factorial designs.

\subsection{Definition of $B_s(d)$ characteristic}
\label{sec:def-Bs}
Suppose there are $m$ controllable factors with two
levels. We represent 
an $n$-run design $d$ by $X(d) \in \{-1,+1\}^{n\times m}$, an $n\times m$
matrix of $-1$'s and $+1$'s. The $(i,j)$th element of $X(d)$,
$x_{ij}(d)$, is the level of the $j$th factor in the $i$th run.
Let $S = \{j_1,\ldots,j_s\} \subseteq \{1,\ldots,m\}$.
Let $\Bx_S(d)$ be the component-wise product of the
$j_1$th, $\ldots$, $j_s$th columns of $X(d)$. The $i$th element of
$\Bx_S(d)$ can be written as $\displaystyle\prod_{j \in S}x_{ij}(d)$.
Note that for any two subsets $S, T\subset \{1,\ldots,m\}$, 
the component-wise product of $\Bx_S(d)$ and $\Bx_T(d)$, say $\Bx_S(d) \odot
\Bx_T(d)$, is written as
$\Bx_S(d) \odot \Bx_T(d) = \Bx_{S \triangle T}(d)$,
where
$S\triangle T = (S\cup T)\setminus (S\cap T)$.
We denote the cardinality of $S \subset \{1,\ldots,m\}$ by $|S|$.
Then $|S| = s$ for $S = \{j_1,\ldots,j_s\}$. 
Define $j_S(d)$ as the sum of all the elements of $\Bx_S(d)$, i.e., 
$j_S(d) = \displaystyle\sum_{i = 1}^n\prod_{j \in S}x_{ij}(d)$.
For $s = 1,\ldots,m$, we define
\[
 B_s(d) = \frac{1}{n^2}\sum_{S:\ |S|=s}(j_S(d))^2.
\]
$\{B_s(d), s = 1,\ldots,m\}$ is the key item in this paper. We call it
$B_s(d)$ characteristic.

\subsection{Generalized minimum aberration and affinely full-dimensional
  factorial designs}
Now we give the relation between $B_s(d)$ characteristic and the
generalized minimum aberration criterion and affinely full-dimensional
factorial designs.

First we note that the set of $j_S(d)$ values over all the possible
$S \subseteq \{1,\ldots,m\}$ has all the information of
the design $d$. In fact, Tang (2001) shows that a design
$d$ is uniquely determined by the set of $j_S(d)$ values. 
Another basic fact is relation between $j_S(d)$ values and the
coefficients
in the indicator function of $d$ defined by Fontana, Pistone and Rogantin
(2000). From the definition of the indicator function, $j_S(d)/n =
b_S/b_{\phi}$ holds, where $b_S$ and $b_{\phi}$ are the 
coefficients of the term corresponding to $S$ and the
constant term, respectively, in the indicator function of 
$d$. See Fontana, Pistone and Rogantin (2000) for detail. 

On the other hand, $B_s(d)$ characteristic has the information of the
aberration of designs. For example, 
if two levels are equireplicated for each factor of the design $d$,
$B_1(d) = 0$ holds. For the orthogonal designs, 
$B_2(d) = 0$ holds. If $d$ is a regular design, $B_3(d)
= 0$ holds for designs of the resolution IV, $B_3(d) =
B_4(d) = 0$ holds for designs of the resolution V, and so on.
Considering these facts and the hierarchical
assumption, Tang and Deng (1999) defined the generalized minimum
aberration criterion as to sequentially minimize $B_1(d), B_2(d),
\ldots, B_m(d)$.

We also give the relation of the $j_S(d)$ values
and the affinely full-dimensional factorial design. 
Note that, for regular designs, each $j_S(d)/n$ is $+1,
-1$ or $0$. By definition, $|j_S(d)/n| = 1$ implies an aliasing
relation, 
whereas $|j_S(d)/n| = 0$ implies an orthogonality. 
For non-regular designs, on the other hand, $|j_S(d)/n|$ can be strictly
between $0$ and $1$, leading to a partial aliasing relation. The 
affinely full-dimensional factorial design can be characterized as the
design satisfying $0 \leq |j_S(d)/n| < 1$ for all $S \subseteq
\{1,\ldots,m\}$. See Lemma 2.2 of Aoki and Takemura (2009) for detail.

From these considerations, the relation between the generalized
minimum aberration criterion and the affinely full-dimensionality is
shown to some extent. Since $B_s(d)$ characteristic is the squared
total of $j_S(d)/n$ for all $S$ satisfying $|S| = s$, minimizing
$B_s(d)$ coincides with  minimizing each $j_S(d)$ for $|S| = s$ 
to some extent. The difference is that the generalized minimum aberration
criterion considers sequentially minimizing $B_1(d),
B_2(d),\ldots,B_m(d)$, whereas the affinely full-dimensionality
considers simultaneous control that each $|j_S(d)/n|$ is strictly less
than $1$. The aim of this paper is to investigate this relation from the
viewpoint of the model-robustness.

\section{Optimal criteria for model-robustness}
\label{sec:optimal-criteria}
To evaluate the model-robustness of the designs, one approach is to
consider the estimation capacity defined by 
Cheng, Steinberg and Sun
(1999). Though the original definition by 
Cheng, Steinberg and Sun (1999) is restricted to the regular designs, 
this concept is generalized by Cheng, Deng and Tang (2002) to non-regular
designs. In this paper, we generalize their works and give general
model-robustness criteria. 

When we choose fractional factorial designs, we can rely on various
optimal criteria such as $D$-optimality based on the information matrix
if the model to be considered is known. On the other hand, if the
model is unknown, which is more realistic situation, we have to
evaluate the model-robustness. In this paper, we consider the situation
where (i) all the main effects are of primary interest and their
estimates are required, (ii) the experimenters suppose that there are $f$
active two-factor interaction effects and $g$ active three-factor
interaction effects, but which of two- and three-factor interactions are
active is unknown and (iii) all the four-factor and higher-order
interactions are negligible. This situation is a natural extension of
the setting of Cheng, Deng and Tang (2002), where the case of $g = 0$
for equireplicated designs. 
Another important case is $\displaystyle f = {m \choose 2}$, meaning
that (i) all the main and the two-factor interaction effects are of interest
and their estimates are required, (ii) there are $g$ active
three-factor interactions, but which of the three-factor interactions
are active is unknown and (iii) all the four-factor and higher-order
interactions are negligible. The aim of our model-robustness criteria is
to evaluate the influence of contamination of active interaction effects
on the parameter estimation.

\subsection{$D_{f,g}$-criterion and $S_{f,g}^2$-criterion}
First we derive an information matrix in our settings. Let 
${\cal P}$ be the set of all the $\displaystyle{m\choose{2}}$ subsets of
the size two of
$\{1,\ldots,m\}$. Similarly, let 
${\cal Q}$ be the set of all the $\displaystyle{m\choose{3}}$ subsets of
the size three of
$\{1,\ldots,m\}$. We have
\[
 {\cal P} = \{\{1,2\},\{1,3\},\ldots,\{m-1,m\}\},
\]
\[
 {\cal Q} = \{\{1,2,3\},\{1,2,4\},\ldots,\{m-2,m-1,m\}\}
\]
and define
\[
\displaystyle |{\cal P}| = {m \choose 2} = F,\ \ 
\displaystyle |{\cal Q}| = {m \choose 3} = G
\]
for later use. Let 
${\cal F}\subset {\cal P}$ and ${\cal G} \subset {\cal Q}$ 
be $f$ active two-factor interactions and $g$ active three-factor
interactions, respectively.
We write $|{\cal F}| = f$ and $|{\cal G}| =
g$. Though we suppose that ${\cal F}$ and ${\cal G}$ are
unknown, it is natural to restrict the models to be considered to satisfy 
the following hierarchical assumption.
\begin{definition}
${\cal F}$ and ${\cal G}$ are called {\it hierarchically consistent} if
\[
 (i_1,i_2,i_3) \in {\cal G} \ \Longrightarrow\ 
 (i_1,i_2), (i_1,i_3), (i_2,i_3) \in {\cal F}.
\]
\end{definition}

For given ${\cal F}$ and ${\cal G}$, we consider a linear model
\[
 \By = \mu\Bone_n + X(d)\Bbeta_1 + Y_{{\cal F}}(d)\Bbeta_2 + 
Z_{{\cal G}}(d)\Bbeta_3 + \Bvarepsilon, 
\]
where $\By$ is the $n\times 1$ vector of observations, 
$\mu$ is an unknown parameter of the general mean, 
$X(d)$ is the $n\times m$ matrix defined in Section \ref{sec:def-Bs}, 
$\Bbeta_1$ is the $m\times 1$ vector of the main effects, 
$Y_{{\cal F}}(d)$ is an $n\times f$ matrix consisting of the $f$ columns 
$\Bx_S(d), S\in {\cal F}$, 
$\Bbeta_2$ is the $f\times 1$ vector of the active two-factor
interactions, $Z_{{\cal G}}(d)$ is an $n\times g$ matrix consisting of
the $g$ columns $\Bx_S(d), S\in {\cal G}$, 
$\Bbeta_3$ is the $g\times 1$ vector of the active three-factor
interactions and 
$\Bvarepsilon$ is an $n\times 1$ random vector satisfying 
$E(\Bvarepsilon) = \Bzero, var(\Bvarepsilon) = \sigma^2 I_n$. Let
$X_{{\cal F}, {\cal G}} = [\Bone_n\ \vdots\ X(d)\ \vdots\
Y_{{\cal F}}(d)\ \vdots\ Z_{{\cal G}}(d)]$.
Then the information matrix for the observations of $d$ is written as
\begin{equation}
\begin{array}{rcl}
 M_{{\cal F}, {\cal G}}(d) & = & \displaystyle\frac{1}{n}X_{{\cal F}, {\cal
 G}}(d)'X_{{\cal F}, {\cal G}}(d)\\
& = &
\left[
{\begin{array}{cccc}
1 & \frac{1}{n}\Bone_n'X(d) & \frac{1}{n}\Bone_n'Y_{{\cal F}}(d) & 
\frac{1}{n}\Bone_n'Z_{{\cal G}}(d)\\
\frac{1}{n}X(d)'\Bone_n & \frac{1}{n}X(d)'X(d) &
 \frac{1}{n}X(d)'Y_{{\cal F}}(d) & 
\frac{1}{n}X(d)'Z_{{\cal G}}(d)\\
\frac{1}{n}Y_{{\cal F}}(d)'\Bone_n & \frac{1}{n}Y_{{\cal F}}(d)'X(d) &
 \frac{1}{n}Y_{{\cal F}}(d)'Y_{{\cal F}}(d) & 
\frac{1}{n}Y_{{\cal F}}(d)'Z_{{\cal G}}(d)\\
\frac{1}{n}Z_{{\cal G}}(d)'\Bone_n & \frac{1}{n}Z_{{\cal G}}(d)'X(d) &
 \frac{1}{n}Z_{{\cal G}}(d)'Y_{{\cal F}}(d) & 
\frac{1}{n}Z_{{\cal G}}(d)'Z_{{\cal G}}(d)
\end{array}}.
\right]
\end{array}
\label{eqn:information-matrix}
\end{equation}
If $\{{\cal F, G}\}$ is known, we can rely on various optimal
criteria based on $M_{{\cal F, G}}(d)$ to choose $d$. For example, 
$D$-optimal criterion is to choose the design that maximize 
$\det M_{{\cal F, G}}(d)$. For the case that
$\{{\cal F, G}\}$ is unknown, it is natural to consider the average
performance over all possible combinations of 
$f$ two-factor interaction effects and $g$ three-factor interaction
effects. To clarify the arguments, we consider probability functions
over the set of all the subsets of, ${\cal P}, {\cal Q}$, i.e., 
$2^{{\cal P}}, 2^{{\cal Q}}$, and consider the expectation of
$\det M_{{\cal F, G}}(d)$ with respect to this probability function.
 If we have no prior information, it is
natural to consider the uniform distribution
\[
 p({\cal F}, {\cal G}) = \left\{\begin{array}{ll}
{\rm Const}, & \mbox{if}\ {\cal F}\ \mbox{and}\ {\cal G}\ \mbox{are
 hierarchically consistent}\\
0, & \mbox{otherwise}.
\end{array}
\right.
\]
Consequently, we can use the expectation $D_{f, g} = E_{p}[\det M_{{\cal
F, G}}(d)]$ to evaluate the model-robustness. We call this a 
$D_{f, g}$-optimal criterion.

However, there is a problem that the calculation of 
$\det M_{{\cal F, G}}(d)$ is difficult. For this problem, we follow the
approach by Cheng, Deng and Tang (2002) and consider minimizing 
$E_{p}[{\rm tr}(M_{{\cal F,G}}(d))^2]$ instead of maximizing $E_p[\det
M_{{\cal F, G}}(d)]$. Note that the calculation of
${\rm tr}(M_{{\cal F,G}}(d))^2$ is considerably easier than that of 
$\det M_{{\cal F, G}}(d)$. It is also known that minimizing 
${\rm tr}(M_{{\cal F,G}}(d))^2$ is a good surrogate for  maximizing 
$\det M_{{\cal F, G}}(d)$. See Cheng (1996) for example.
In addition, since all the diagonal elements of $M_{{\cal F,
G}}(d)$ are $1$, minimizing $E_p[{\rm tr}(M_{{\cal F,G}}(d))^2]$ is
equivalent to minimizing the expectation of 
the sum of squares of all the off-diagonal elements of $M_{{\cal
F, G}}(d)$. We write this value as
\[
 S_{f,g}^2 = E_p[\mbox{sum of squares of all the off-diagonal elements
 of}\ M_{{\cal F, G}}(d)]
\]
and define our criterion.
\begin{definition}
$S_{f, g}^2$-optimal criterion is to choose designs that minimize
 $S_{f, g}^2$.
\end{definition}

\subsection{Calculation of $S_{f,g}^2$ values}
To evaluate the $S_{f, g}^2$ value, we have to calculate all the
off-diagonal elements of $M_{{\cal F, G}}(d)$. 
We consider each block in the
partitioned matrix (\ref{eqn:information-matrix}) separately.
First we see that the sum of squares of all the elements of 
$(1/n)\Bone_n'X(d)$ is $\displaystyle \sum_{i = 1}^m
      \displaystyle\frac{(j_{\{i\}}(d))^2}{n^2} = B_1(d)$ by definition.
Similarly, the sum of squares of all the off-diagonal elements of 
$(1/n)X(d)'X(d)$ is $\displaystyle
      2\sum_{S \in {\cal P}}\frac{(j_{S}(d))^2}{n^2} = 2B_2(d)$ by definition.
Since the calculations of all the other blocks depend on the probability
function $p({\cal F, G})$, we have the following expression.
\begin{equation}
\begin{array}{rl}
 S_{f,g}^2 = & 2B_1(d) + 2B_2(d) +
  2E_p\left[\displaystyle\frac{1}{n^2}\sum_{S \in
 {\cal F}}(j_S(d))^2\right] +
  2E_p\left[\displaystyle\frac{1}{n^2}\sum_{i = 1}^m\sum_{S \in {\cal
      F}}(j_{\{i\}\triangle S}(d))^2
\right]\\
& + E_p\left[
\displaystyle\frac{1}{n^2}\mathop{\sum\sum}_{\stackrel{S,T
      \in {\cal F}}{S \neq T}}(j_{S\triangle T}(d))^2\right] +
2E_p\left[
      \displaystyle\frac{1}{n^2}\sum_{S \in {\cal G}}(j_S(d))^2
\right]\\
& + 2E_p\left[
      \displaystyle\frac{1}{n^2}\sum_{i = 1}^m\sum_{S \in {\cal
      G}}(j_{\{i\}\triangle S}(d))^2
\right]
+ 2E_p\left[
      \displaystyle\frac{1}{n^2}\sum_{S \in {\cal F}}\sum_{T \in {\cal
      G}}(j_{S\triangle T}(d))^2
\right]\\
& + E_p\left[
\displaystyle\frac{1}{n^2}\mathop{\sum\sum}_{\stackrel{S,T
      \in {\cal G}}{S \neq T}}(j_{S\triangle T}(d))^2
\right]
\end{array}
\label{eqn:s_fg^2}
\end{equation}

Now all we have to do is to evaluate 
the expectations of (\ref{eqn:s_fg^2}) for specific
values of $f$, $g$ and $p({\cal F, G})$. In this paper, we only consider
the cases that $p({\cal F, G})$ is the uniform distribution on the
symmetric support for the factors $\{1,\ldots,m\}$. For these cases,
$S_{f,g}^2$ is expressed as a linear combination of
$B_1(d),B_2(d),\ldots,B_6(d)$. Note that $B_6(d)$ only arises in the last
term of (\ref{eqn:s_fg^2}) as the contribution of $(j_{S\triangle
T}(d))^2$ where $S$ and $T$ are disjoint. Though the uniform assumption
on the symmetric support is natural, there are various important
situations where the support of $p({\cal F, G})$ is asymmetric. For this
point, we consider shortly in Section \ref{sec:discussion}.

Unfortunately, it seems very difficult to derive $S_{f,g}^2$
values for general $f, g$ values. One of the simpler problems,
evaluation of $S_{f, 1}^2$, is also difficult. In this paper, we obtain
the results on some specific cases.

\subsubsection{Calculation of $S_{f,0}^2$}
First we consider the situation that all the three-factor interaction
effects are negligible. This situation is considered in Cheng, Deng and Tang
(2002) for equireplicated designs and therefore our result is an
extension of their result. In this case, the relation 
(\ref{eqn:s_fg^2}) becomes
\begin{equation}
\begin{array}{rl}
 S_{f,0}^2 = & 2B_1(d) + 2B_2(d) +
  2E_p\left[\displaystyle\frac{1}{n^2}\sum_{S \in
 {\cal F}}(j_S(d))^2\right] +
  2E_p\left[\displaystyle\frac{1}{n^2}\sum_{i = 1}^m\sum_{S \in {\cal
      F}}(j_{\{i\}\triangle S}(d))^2
\right]\\
& + E_p\left[
\displaystyle\frac{1}{n^2}\mathop{\sum\sum}_{\stackrel{S,T
      \in {\cal F}}{S \neq T}}(j_{S\triangle T}(d))^2\right],
\end{array}
\label{eqn:S_f0^2}
\end{equation}
and we consider the uniform distribution on ${\cal P}$, 
\[
 p({\cal F}) = 
\displaystyle\frac{1}{\displaystyle{F \choose{f}}}.
\]
The result is summarized as follows.
\begin{theorem}
\label{thm:S_f0}
$S_{f, 0}^2$ is written as $S_{f,0}^2 = \displaystyle\sum_{s =
 1}^4a_sB_s(d)$, where
\[\begin{array}{l}
a_1 = 2\left(1 + \displaystyle\frac{f(m-1)}{F}\right),\ 
a_2 = 2\left(1 + \displaystyle\frac{f}{F} +
	\frac{f(f-1)}{F(F-1)}(m-2)\right),\\ 
a_3 = \displaystyle\frac{6f}{F}\ \mbox{and}\ 
a_4 = \displaystyle\frac{6f(f-1)}{F(F-1)}.
\end{array}
\]
\end{theorem}
We give the proof in Appendix A. 
Note that the result except for $a_1$ is also given in Cheng, Deng and
Tang (2002). 
From Theorem \ref{thm:S_f0}, we see the following result.
\begin{proposition}
\label{prop:S_f0}
The relation $a_1 > a_2 > a_3 > a_4$ holds for $m > 3$ for Theorem
 \ref{thm:S_f0}.
\end{proposition}
We give the proof in Appendix B. Proposition \ref{prop:S_f0}
implies a consistency of the
$S_{f,0}^2$-criterion and the generalized minimum aberration criterion.
We see the optimal designs for these two criteria can be reversed by
empirical studies for $12$-run designs of $5$ factors
in Section \ref{sec:12-run-studies}.

\subsubsection{Calculation of $S_{F,g}^2$}
Next we consider the situation that all the two-factor interactions are
active, i.e., the case of $f = F$. In this case, since 
${\cal F} = {\cal P}$, we consider the uniform distribution on ${\cal
Q}$ as
\[
 p({\cal G}) = \frac{1}{\displaystyle{G \choose{g}}}.
\]
The result is summarized as follows.
\begin{theorem}
\label{thm:S_Fg}
$S_{F, g}^2$ is written as
$S_{F,g}^2 = \displaystyle\sum_{s = 1}^6a_sB_s(d)$, where
\[
\begin{array}{l}
a_1 = 2m + \displaystyle\frac{g(m-1)(m-2)}{G},\\
a_2 = 2m + \displaystyle\frac{2g(m-2)}{G} + \frac{g(g-1)(m-2)(m-3)}{G(G-1)},\\
a_3 = 6 + \displaystyle\frac{2g}{G} + \frac{6g(m-3)}{G},\\
a_4 = 6 + \displaystyle\frac{8g}{G} + \frac{6g(g-1)(m-4)}{G(G-1)},\\
a_5 = \displaystyle\frac{20g}{G}\ \mbox{and}\\
a_6 = \displaystyle\frac{20g(g-1)}{G(G-1)}.
\end{array}
\]
\end{theorem}
We give the proof in Appendix C.
From Theorem \ref{thm:S_Fg}, we have the following result.
\begin{proposition}
\label{prop:S_Fg}
The relation $a_1 > a_2 > a_3 > a_4 > a_5 > a_6$ holds for $m > 5$ for
 Theorem \ref{thm:S_Fg}.
\end{proposition}
We give the proof in Appendix D. Proposition \ref{prop:S_Fg}
implies a consistency of the
$S_{F,g}^2$-criterion and the generalized minimum aberration criterion.

\subsubsection{Calculation of $S_{3,1}^2$}
Next we calculate $S_{3,1}^2$, which means the situation that there are
one active three-factor interaction and three active two-factor
interactions included in the three-factor interaction hierarchically. In
this case, the joint probability function and its marginal probability
functions are written as
\[
 p({\cal F, G}) = 
\left\{\begin{array}{ll}
\displaystyle\frac{1}{G}, & \mbox{if ${\cal F}$ and ${\cal G}$ are
 hierarchically consistent},\\
0, & \mbox{otherwise},
\end{array}
\right.
\]
\[
 p({\cal G}) = \displaystyle\frac{1}{G}
\]
and
\[
p({\cal F}) = \left\{\begin{array}{ll}
\displaystyle\frac{1}{G}, & \mbox{if there exists ${\cal G}$ such that
 ${\cal F}$ and ${\cal G}$ are
 hierarchically consistent}\\
0, & \mbox{otherwise}.
\end{array}
\right.
\]
The result is summarized as follows.
\begin{theorem}
\label{thm:S_31}
$S_{3, 1}^2$ is written as
$S_{3,1}^2 = \displaystyle\sum_{s = 1}^4a_sB_s(d)$, where
\[
\begin{array}{l}
%a_1 = 2\left(1 + \displaystyle\frac{(m-1)(m-2)}{G} + \frac{3}{m}\right),\\
a_1 = 2\left(1 + \displaystyle\frac{9}{m}\right),\\
a_2 = 2\left((m-1) + \displaystyle\frac{4(m-2)}{G}\right),\\
a_3 = \displaystyle\frac{2(3m-5)}{G}\ \mbox{and}\\
a_4 = \displaystyle\frac{8}{G}.
\end{array}
\]
\end{theorem}
We give the proof in Appendix E.
From Theorem \ref{thm:S_31}, we have the following result.
\begin{proposition}
\label{prop:S_31}
The relation $a_2 > a_1 > a_3 > a_4$ holds for $m > 3$ for
 Theorem \ref{thm:S_31}.
\end{proposition}
We give the proof in Appendix F. 
Proposition \ref{prop:S_31} shows quite different tendency against 
Proposition \ref{prop:S_f0} and Proposition \ref{prop:S_Fg}, i.e., 
$a_1 < a_2$ holds. This fact implies an
essential difference between the $S_{3,1}^2$-criterion and the
generalized minimum aberration criterion, i.e., the $S_{3,1}^2$-criterion
puts more importance on the orthogonality between the columns of $X(d)$
than the equireplicateness of two levels, which is mostly emphasized in
the generalized minimum aberration criterion. Consequently, we can suppose the
optimal designs for two criteria can be reversed. We 
investigate this point by empirical studies for $12$-run designs of
$5$ factors in Section \ref{sec:12-run-studies}.

\subsection{$S_{f,g}^2$-optimal designs for $12$-run designs}
\label{sec:12-run-studies}
To clarify the relation between the $S_{f,g}^2$-criterion and the
generalized minimum aberration, we consider fractional factorial
$12$-run designs of $5$ factors. We are also interested in the affinely
full-dimensionality of the optimal designs. Note that all the fractional
factorial designs with $n > 2^{m-1}$ are affinely full-dimensional since
these designs cannot be a proper subset of any regular fractional
factorial designs. See Aoki and Takemura (2009) for detail. 
Another reason that we consider $12$-run designs is related to the existence
of Hadamard matrix of order $12$. 
Since the run size $n = 12$ is even, it is clear that the
generalized minimum aberration criterion prefers the designs with 
equireplicated levels. It is also clear that we can easily construct 
orthogonal designs by choosing the columns of Hadamard matrices of order
$12$. See Deng, Li and Tang (2000) for example. From these
considerations, we see that the optimal designs with the generalized
minimum aberration satisfy $B_1(d) = B_2(d) = 0$. In fact, all the
$12\times 5$ designs constructed from five columns (except for
$\Bone_{12}$) of Hadamard matrices
of order $12$, say $d_h$, satisfy
\[
 B_1(d_h) = B_2(d_h) = 0,\ \ B_3(d_h) = 1.1111,\ \ B_4(d_h) = 0.5556.
\]
We compare the $B_s(d)$ characteristics of the $S_{f,g}^2$-optimal
designs with this value.

We enumerate all the fractional
factorial designs of $5$ factors with $12$ runs and obtain
$S_{f,0}^2$-optimal designs for $f = 1,\ldots,5$ and $S_{3,1}^2$-optimal
design. We have confirmed that all the optimal designs are equivalent to
the design shown in Table \ref{tbl:d_s} 
by permuting factors or levels and changing signs. 
\begin{table*}[htbp]
\begin{center}
\caption{$S_{f,0}^2$- and $S_{3,1}^2$-optimal $12$-run
 design of $5$ factors}
\label{tbl:d_s}
\begin{tabular}{|rrrrr|}\hline
$1$ & $1$ & $1$ & $1$ & $1$\\
$1$ & $1$ & $-1$ & $-1$ & $-1$\\
$1$ & $-1$ & $1$ & $1$ & $-1$\\
$1$ & $-1$ & $1$ & $-1$ & $1$\\
$1$ & $-1$ & $-1$ & $1$ & $1$\\
$-1$ & $1$ & $1$ & $1$ & $-1$\\
$-1$ & $1$ & $1$ & $-1$ & $1$\\
$-1$ & $1$ & $-1$ & $1$ & $1$\\
$-1$ & $-1$ & $1$ & $1$ & $1$\\
$-1$ & $-1$ & $1$ & $-1$ & $-1$\\
$-1$ & $-1$ & $-1$ & $1$ & $-1$\\
$-1$ & $-1$ & $-1$ & $-1$ & $1$\\ \hline
\end{tabular}
\end{center}
\end{table*}
This design satisfies the 
$S_{f,0}^2$-, $f = 1,\ldots,5$ and $S_{3,1}^2$-optimality
simultaneously. 
The $B_s(d)$ characteristics for this design, say $d_s$, are 
\[
 B_1(d_s) = 0.138889,\ B_2(d_s) = 0,\ B_3(d_s) = 0.27778,\ B_4(d_s) =
 0.5556.
\]
Since $B_1(d_s) > B_1(d_h)$, $d_s$ does not have the generalized minimum
aberration. We also see that $d_s$ is an orthogonal design and 
$B_4(d_s) = B_4(d_h)$. The difference between the two designs in view of
$B_s(d)$ characteristics lies in $B_1(d)$ and $B_3(d)$. We see that the
generalized minimum aberration criterion puts the importance on $B_1(d)$,
whereas the $S_{f,g}^2$-criteria consider the overall values. 
Table \ref{tbl:opt-summary} shows the $S_{f,0}^2, f = 1,\ldots,5$ and
$S_{3,1}^2$ values for $d_h$ and $d_s$. 
\begin{table*}[htbp]
\begin{center}
\caption{$S_{f,0}^2, f = 1,\ldots,5$ and $S_{3,1}^2$ values for two
 deigns, $d_h$ and $d_s$}
\label{tbl:opt-summary}
\begin{tabular}{ccccccc}\hline
& $S_{1,0}^2$ & $S_{2,0}^2$ & $S_{3,0}^2$ & $S_{4,0}^2$ & $S_{5,0}^2$ &
 $S_{3,1}^2$\\ \hline
$d_s$ & $0.5556$ & $0.9074$ & $1.3333$ & $1.8333$ & $2.4074$ & $1.7778$\\
$d_h$ & $0.6667$ & $1.4074$ & $2.2222$ & $3.1111$ & $4.0741$ & $2.6667$\\ \hline
\end{tabular}
\end{center}
\end{table*}

We see that both $d_h$ and $d_s$ are affinely full-dimensional, and
therefore not proper subsets of any regular fractional factorial
designs. This fact implies that the simple strategies such as choosing
$12$ rows from regular $2^{5-1}$ fractional factorial designs to
construct a $12$-run design can cause a design of bad performance, in
view of the generalized minimum aberration and model-robustness.

\section{Discussion}
\label{sec:discussion}
We propose a general method to evaluate model-robustness for non-regular
two-level designs. Though we suppose, in this paper, the four- and
higher-factor
interactions are negligible, which is considered to be a natural
assumption in actual situations, we can easily generalize our method to
incorporate higher-factor interactions. 

It is also possible to calculate $S_{f,g}^2$ values for small $f,g$ such as
$S_{4,1}^2$, $S_{5,1}^2$ or $S_{5,2}^2$. Though the calculations will be
rather complicated, they are indeed based on a simple counting. It is
true that the 
assumption that the experimenters only have an information on the
number of the interactions in the true model seems unnatural in actual
situations. However, we think that the $S_{f,g}^2$ values for small
$f,g$ can be used to evaluate the model-robustness. Here we regard $f$
and $g$ as the degree of contamination of interactions.

Though we only consider the cases that $p({\cal F, G})$ is 
the uniform distribution on the
symmetric support for the factors $\{1,\ldots,m\}$, there are 
various important
situations where the support of $p({\cal F, G})$ is asymmetric. 
One of the examples for asymmetric cases is that (i) there are $m_1$
controllable factors and $m - m_1$ noise factors, (ii) all the main
effects and two-factor interaction effects between the controllable
factor and the noise factor are of primary interest and their
estimates are required, (iii) all the two-factor interactions between
two controllable factors are negligible. all the three- and
higher-factor interactions are also negligible, and (iv) among the
two-factor interactions between two noise factors, there are $f - m_1(m
- m_1)$ active interactions. For this situation, it is the important problem
to investigate the model-robustness of designs for the contamination of
the two-factor interactions between two noise factors. However, for such
asymmetric situation, $S_{f,g}^2$ values cannot be expressed as a
linear combination of $B_s(d)$ characteristic. We postpone this
attractive topic to future works.

\section*{Appendix A. Proof of Theorem \ref{thm:S_f0}}
We evaluate the terms of (\ref{eqn:S_f0^2}) separately.
First we have
\[
E_p\left[\displaystyle\frac{1}{n^2}\sum_{S \in
 {\cal F}}(j_S(d))^2\right] = 
\displaystyle\frac{1}{\displaystyle{F
 \choose{f}}}\sum_{{\cal F}\subset {\cal P}}\sum_{S \in {\cal
 F}}\left(\frac{j_S(d)}{n}\right)^2 = 
\displaystyle\frac{1}{\displaystyle{F
 \choose{f}}}\frac{\displaystyle{F \choose{f}}f}{F}\sum_{S \in {\cal
 P}}\left(\frac{j_S(d)}{n}\right)^2 = \frac{f}{F}B_2(d).
\]
Next from 
\begin{equation}
 \{i\}\triangle S = \left\{\begin{array}{ll}
S\setminus i, & \mbox{if}\ i \in S,\\
\{i, S\}, & \mbox{otherwise}
\end{array}
\right.
\label{eqn:i-triangle-S}
\end{equation}
for $S \in {\cal F}$, we have
\[\begin{array}{l}
E_p\left[\displaystyle\frac{1}{n^2}\sum_{i = 1}^m\sum_{S \in {\cal
      F}}(j_{\{i\}\triangle S}(d))^2
\right] = \displaystyle\frac{1}{\displaystyle{F \choose{f}}}\sum_{{\cal
      F}\subset
      {\cal P}}\sum_{i = 1}^m\sum_{S \in {\cal
      F}}\left(\frac{j_{\{i\}\triangle S}(d)}{n}\right)^2 = 
\frac{f}{F}\sum_{i = 1}^m\sum_{S \in {\cal
      P}}\left(\frac{j_{\{i\}\triangle S}(d)}{n}\right)^2\\
\hspace*{6mm} = \displaystyle\frac{f}{F}\left((m-1)\sum_{i =
			    1}^m
			    \left(\displaystyle\frac{j_{\{i\}}(d)}{n}\right)^2
+ 3\sum_{S \in {\cal Q}}\left(\displaystyle\frac{j_{S}(d)}{n}\right)^2
\right) = \displaystyle\frac{f}{F}((m-1)B_1(d) + 3B_3(d)).
\end{array}
\]
Similarly, for distinct $i,j,k,\ell \in \{1,\ldots,m\}$ we have
\begin{equation}
 S \triangle T = \left\{\begin{array}{ll}
\{i,j\}, & \mbox{for}\ S=\{i,k\}, T = \{j,k\},\\
\{i,j,k,\ell\}, & \mbox{for}\ S=\{i,j\}, T = \{k,\ell\}.
\end{array}
\right.
\label{eqn:s-triangle-t-in-F}
\end{equation}
Then it follows
\[\begin{array}{l}
 E_p\left[
\displaystyle\frac{1}{n^2}\mathop{\sum\sum}_{\stackrel{S,T
      \in {\cal F}}{S \neq T}}(j_{S\triangle T}(d))^2\right] =
 \displaystyle\frac{f(f-1)}{F(F-1)}\mathop{\sum\sum}_{\stackrel{S,T
      \in {\cal P}}{S \neq T}}\left(\frac{j_{S\triangle
			       T}(d)}{n}\right)^2\\
\hspace*{6mm} = 
\displaystyle\frac{f(f-1)}{F(F-1)}(2(m-2)B_2(d) + 6B_4(d))
\end{array}
\]
by simple counting. 
From the above calculations, we have the
theorem. \hspace*{\fill}Q.E.D.

\section*{Appendix B. Proof of Proposition \ref{prop:S_f0}}
For $m > 3$, we have
\[
 a_1 - a_2 = \frac{2f(m-2)(F-f)}{F(F-1)} > 0
\]
and
\[
 a_3 - a_4 = \frac{6f(F-f)}{F(F-1)} > 0.
\]
From the relations
\[
 a_2 - a_3 = \frac{2}{F(F-1)}\{(m-2)f^2 - (2F+m-4)f + F(F-1)\}
\]
and 
\[
 (2F + m - 4)^2 - 4F(F-1)(m-2)  = -(m-2)^2(m^3 - m^2 - 5m - 4) < 0,
\]
we have $a_2 > a_3$ for all $f$. 
Therefore we have shown the proposition. \hspace*{\fill}Q.E.D.

\section*{Appendix C. Proof of Theorem \ref{thm:S_Fg}}
From ${\cal F} = {\cal P}$ and simple counting, we have 
\[
  2E_p\left[\displaystyle\frac{1}{n^2}\sum_{S \in
 {\cal F}}(j_S(d))^2\right] = 
\displaystyle\frac{2}{n^2}\sum_{S \in
 {\cal P}}(j_S(d))^2 = 2B_2(d), 
\]
\[
  2E_p\left[\displaystyle\frac{1}{n^2}\sum_{i = 1}^m\sum_{S \in {\cal
      F}}(j_{\{i\}\triangle S}(d))^2
\right] = 
  \displaystyle\frac{2}{n^2}\sum_{i = 1}^m\sum_{S \in {\cal
      P}}(j_{\{i\}\triangle S}(d))^2 = 2((m-1)B_1 + 3B_3(d))
 \]
and
\[
E_p\left[
\displaystyle\frac{1}{n^2}\mathop{\sum\sum}_{\stackrel{S,T
      \in {\cal F}}{S \neq T}}(j_{S\triangle T}(d))^2\right] = 
\displaystyle\frac{1}{n^2}\mathop{\sum\sum}_{\stackrel{S,T
      \in {\cal P}}{S \neq T}}(j_{S\triangle T}(d))^2 = 
2(m-2)B_2(d) + 6B_4(d).
\]
Therefore (\ref{eqn:s_fg^2}) becomes
\[
\begin{array}{rcl}
 S_{f,g}^2 
%& = & 2B_1(d) + 2B_2(d) +
%  2B_2(d) +
%  2((m-1)B_1 + 3B_3(d))\\
%& & + 2(m-2)B_2(d) + 6B_4(d) + 
%2E_p\left[
%      \displaystyle\frac{1}{n^2}\sum_{S \in {\cal G}}(j_S(d))^2
%\right] + 
%2E_p\left[
%      \displaystyle\frac{1}{n^2}\sum_{i = 1}^m\sum_{S \in {\cal
%      G}}(j_{\{i\}\triangle S}(d))^2
%\right]\\
%& & + 2E_p\left[
%      \displaystyle\frac{1}{n^2}\sum_{S \in {\cal F}}\sum_{T \in {\cal
%      G}}(j_{S\triangle T}(d))^2
%\right]
%+ E_p\left[
%\displaystyle\frac{1}{n^2}\mathop{\sum\sum}_{\stackrel{S,T
%      \in {\cal G}}{S \neq T}}(j_{S\triangle T}(d))^2
%\right]\\
& = & 2mB_1(d) + 2mB_2(d) + 6B_3(d) + 6B_4(d) + 
2E_p\left[
      \displaystyle\frac{1}{n^2}\sum_{S \in {\cal G}}(j_S(d))^2
\right]\\
& & +  
2E_p\left[
      \displaystyle\frac{1}{n^2}\sum_{i = 1}^m\sum_{S \in {\cal
      G}}(j_{\{i\}\triangle S}(d))^2
\right]
+ 2E_p\left[
      \displaystyle\frac{1}{n^2}\sum_{S \in {\cal F}}\sum_{T \in {\cal
      G}}(j_{S\triangle T}(d))^2
\right]\\
& & + E_p\left[
\displaystyle\frac{1}{n^2}\mathop{\sum\sum}_{\stackrel{S,T
      \in {\cal G}}{S \neq T}}(j_{S\triangle T}(d))^2
\right].
\end{array}
\]
Now we consider the expectations above separately. 
From simple counting, we have 
\[
E_p\left[
      \displaystyle\frac{1}{n^2}\sum_{S \in {\cal G}}(j_S(d))^2
\right] = \displaystyle\frac{g}{G}\sum_{S \in {\cal
      Q}}\left(\frac{j_S(d)}{n}\right)^2 = \frac{g}{G}B_3(d), 
\]
\[
E_p\left[
      \displaystyle\frac{1}{n^2}\sum_{i = 1}^m\sum_{S \in {\cal
      G}}(j_{\{i\}\triangle S}(d))^2
\right] = \displaystyle\frac{g}{G} \sum_{i = 1}^m\sum_{S \in {\cal
      Q}}\left(\frac{j_{\{i\}\triangle S}(d)}{n}\right)^2 =
      \displaystyle\frac{g}{G}((m - 2)B_2(d) + 4B_4(d))
\]
from (\ref{eqn:i-triangle-S}) for $S \in {\cal Q}$, 
\[\begin{array}{l}
E_p\left[
      \displaystyle\frac{1}{n^2}\sum_{S \in {\cal F}}\sum_{T \in {\cal
      G}}(j_{S\triangle T}(d))^2
\right] = \displaystyle\frac{g}{G}\sum_{S \in {\cal P}}\sum_{T \in {\cal
      Q}}\left(\frac{j_{S\triangle T}(d)}{n}\right)^2\\
=  \displaystyle\frac{g}{G}\left(
\frac{(m-1)(m-2)}{2}B_1(d) + 3(m-3)B_3(d) + 10B_5(d)
\right)
\end{array}\]
from
\[
 S \triangle T = \left\{\begin{array}{ll}
\{i_1\}, & \mbox{for}\ S = \{i_2,i_3\},\ T = \{i_1,i_2,i_3\}\\
\{i_1,i_2,i_3\}, & \mbox{for}\ S = \{i_3,i_4\},\ T = \{i_1,i_2,i_4\}\\
\{i_1,i_2,i_3,i_4,i_5\}, & \mbox{for}\ S = \{i_4,i_5\},\ T = \{i_1,i_2,i_3\}
\end{array}
\right.
\]
for distinct $i_1,\ldots,i_5 \in \{1,\ldots,m\}$ and
\[\begin{array}{l}
E_p\left[
\displaystyle\frac{1}{n^2}\mathop{\sum\sum}_{\stackrel{S,T
      \in {\cal G}}{S \neq T}}(j_{S\triangle T}(d))^2
\right] = \displaystyle\frac{g(g-1)}{G(G-1)}\mathop{\sum\sum}_{\stackrel{S,T
      \in {\cal Q}}{S \neq T}}\left(\frac{j_{S\triangle
			       T}(d)}{n}\right)^2\\
= \displaystyle\frac{g(g-1)}{G(G-1)}((m-2)(m-3)B_2(d) +
 6(m-4)B_4(d) + 20B_6(d))
\end{array}\]
from
\[
 S \triangle T = \left\{\begin{array}{ll}
\{i_1,i_2\}, & \mbox{for}\ S = \{i_1,i_3,i_4\},\ T = \{i_2,i_3,i_4\}\\
\{i_1,i_2,i_3,i_4\}, & \mbox{for}\ S = \{i_1,i_2,i_5\},\ T = \{i_3,i_4,i_5\}\\
\{i_1,i_2,i_3,i_4,i_5,i_6\}, & \mbox{for}\ S = \{i_1,i_2,i_3\},\ T =
 \{i_4,i_5,i_6\}
\end{array}
\right.
\]
for distinct $i_1,\ldots,i_6 \in \{1,\ldots,m\}$.
From the above calculations, we have the
theorem. \hspace*{\fill}Q.E.D.

\section*{Appendix D. Proof of Proposition \ref{prop:S_Fg}}
For $m > 5$, we have
\[
 a_1 - a_2 = \frac{g(m-2)(m-3)(G-g)}{G(G-1)} > 0,
\]
\[
 a_3 - a_4 = \frac{6g(m-4)(G-g)}{G(G-1)} > 0
\]
and
\[
 a_5 - a_6 = \frac{20g(G-g)}{G(G-1)} > 0.
\]
For the relation between $a_2$ and $a_3$, we have
\[
 a_2 - a_3 = \frac{m-3}{G(G-1)}\{(m-2)g^2 - (4G+m-6)g + 2G(G-1)\}
\]
and
\[\begin{array}{cl}
& (4G + m - 6)^2 - 8G(G-1)(m-2)\\
= & -8(m-4)G^2 + 16(m-4)G + (m-6)^2\\
< & -(m-4)(8G(G-2) - (m-4))\\
< & -(m-4)\left(8\displaystyle\frac{m^2(m-4)}{6}(G-2) - (m-4)\right)\\
= & -\displaystyle\frac{(m-4)^2}{3}(4m^2(G-2) - 3))\\
< & -\displaystyle\frac{(m-4)^2}{3}(4\cdot 5^2\cdot 8 - 3)) < 0
\end{array}\]
since $G > m^2(m-4)/6$ holds for $m > 5$ and $m^2(G-2)$ is a monotone
increasing function of $m$. Similarly, for the relation between $a_4$
and $a_5$, we have
\[
 a_4 - a_5 = \frac{6}{G(G-1)}\{(m-4)g^2 - (2G+m-6)g + G(G-1)\}
\]
and 
\[\begin{array}{cl}
& (2G+m-6)^2 - 4G(G-1)(m-4)\\
= & -4(m-5)G^2 + 8(m-5)G + (m-6)^2\\
< & -(m-5)(4G(G-2) - (m-5)) \\
< & -(m-5)\left(4\displaystyle\frac{m^2(m-5)}{6}(G-2) - (m-5)\right)\\
= & -\displaystyle\frac{(m-5)^2}{3}(2m^2(G-2) - 3)\\
< & -\displaystyle\frac{(m-5)^2}{3}(2\cdot 5^2\cdot 8 - 3) < 0\\
\end{array}
\]
since $G > m^2(m-5)/6$ holds for $m > 5$ and $m^2(G-2)$ is a monotone
increasing function of $m$. Therefore we have shown the
proposition. \hspace*{\fill}Q.E.D.

\section*{Appendix E. Proof of Theorem \ref{thm:S_31}}
In this case, we write ${\cal G} = \{U\} \in {\cal Q}$ and 
$Z_{{\cal G}}(d) = \Bx_U(d)$. 
Then (\ref{eqn:s_fg^2}) becomes
\begin{equation}
\begin{array}{rl}
 S_{3,1}^2 = & 2B_1(d) + 2B_2(d) +
  2E_p\left[\displaystyle\frac{1}{n^2}\sum_{S \in
 {\cal F}}(j_S(d))^2\right] +
  2E_p\left[\displaystyle\frac{1}{n^2}\sum_{i = 1}^m\sum_{S \in {\cal
      F}}(j_{\{i\}\triangle S}(d))^2
\right]\\
& + E_p\left[
\displaystyle\frac{1}{n^2}\mathop{\sum\sum}_{\stackrel{S,T
      \in {\cal F}}{S \neq T}}(j_{S\triangle T}(d))^2\right] +
2E_p\left[
      \displaystyle\left(\frac{j_U(d)}{n}\right)^2
\right] + 
2E_p\left[
      \displaystyle\frac{1}{n^2}\sum_{i = 1}^m(j_{\{i\}\triangle U}(d))^2
\right]\\
& + 2E_p\left[
      \displaystyle\frac{1}{n^2}\sum_{S \in {\cal F}}(j_{S\triangle U}(d))^2
\right].
\end{array}
\label{eqn:s_31^2-2}
\end{equation}
We consider all the terms of (\ref{eqn:s_31^2-2}) separately. From
simple counting, we have
\[
  2E_p\left[\displaystyle\frac{1}{n^2}\sum_{S \in
 {\cal F}}(j_S(d))^2\right] = 2\frac{m-2}{G}\sum_{S \in {\cal
 P}}\left(\frac{j_S(d)}{n}\right)^2 = \frac{2(m-2)}{G}B_2(d),
\]
\[\begin{array}{cl}
&  2E_p\left[\displaystyle\frac{1}{n^2}\sum_{i = 1}^m\sum_{S \in {\cal
      F}}(j_{\{i\}\triangle S}(d))^2
\right]\\
  = & \displaystyle\frac{2}{G}\left(
2{{m-1}\choose{2}}\sum_{i = 1}^m\left(\frac{j_{\{i\}}(d)}{n}
\right)^2 + (3+3(m-3))\sum_{S \in {\cal Q}}\left(\frac{j_S(d)}{n}
\right)^2
\right)\\
 = & \displaystyle\frac{2(m-1)(m-2)}{G}B_1(d) + \frac{6(m-2)}{G}B_3(d)
\end{array}
\]
from (\ref{eqn:i-triangle-S}) for $S \in {\cal F}$, 
\[
 E_p\left[
\displaystyle\frac{1}{n^2}\mathop{\sum\sum}_{\stackrel{S,T
      \in {\cal F}}{S \neq T}}(j_{S\triangle T}(d))^2\right] =
      2(m-2)\sum_{S \in {\cal P}}\left(\frac{j_S(d)}{n}
\right)^2 = 2(m-2)B_2(d)
\]
from (\ref{eqn:s-triangle-t-in-F}) where $S \cap T \neq \emptyset$, 
\[
2E_p\left[
      \displaystyle\left(\frac{j_U(d)}{n}\right)^2
\right] = \displaystyle\frac{2}{G}\sum_{U \in {\cal Q}}
      \displaystyle\left(\frac{j_U(d)}{n}\right)^2 = \frac{2}{G}B_3(d),
\]
\[\begin{array}{cl}
2E_p\left[
      \displaystyle\frac{1}{n^2}\sum_{i = 1}^m(j_{\{i\}\triangle U}(d))^2
\right] & = \displaystyle\frac{2}{G}\left((m-2)\sum_{S \in {\cal P}}\left(
\frac{j_S(d)}{n}
\right)^2 
+ 4\sum_{S: |S| = 4}\left(
\frac{j_S(d)}{n}
\right)^2 
\right)\\
& = \displaystyle\frac{2(m-2)}{G}B_2(d) + \frac{8}{G}B_4(d)
\end{array}
\]
from (\ref{eqn:i-triangle-S}) for $S = U \in {\cal Q}$ and
\[
  2E_p\left[
      \displaystyle\frac{1}{n^2}\sum_{S \in {\cal F}}(j_{S\triangle U}(d))^2
\right] = \frac{2}{G}\frac{3G}{m}\sum_{i = 1}^m\left(\frac{j_{\{i\}}(d)}{n}
\right)^2 = \frac{6}{m}B_1(d)
\]
from $S \subset U$ and $S\triangle U = U\setminus S$.
From the above calculations, we have the theorem.\hspace*{\fill}Q.E.D.

\section*{Appendix F. Proof of Proposition \ref{prop:S_31}}
For $m > 3$, we have
\[\begin{array}{cl}
 a_2 - a_1 & = \displaystyle\frac{2}{mG}((m^2 - 2m - 9)G + 4m(m-2))\\
& = \displaystyle\frac{m-2}{3G}(m^3 - 3m^2 - 7m + 33) > 0,
\end{array}\]
\[\begin{array}{cl}
 a_1 - a_3 & = \displaystyle\frac{2}{mG}((m+9)G - m(3m-5))\\
& = \displaystyle\frac{1}{3G}(m^3 + 6m^2 - 28m + 23) > 0
\end{array}\]
and
\[
 a_3 - a_4 = \frac{2}{G}(3m - 9) > 0.
\]
Therefore we have shown the proposition. \hspace*{\fill}Q.E.D.

\end{document}